\documentclass{elsart}
\usepackage{ifpdf}
\usepackage{graphicx,amssymb,lineno}
\ifpdf
\usepackage[%
  pdftitle={Instructions for use of the document class
    elsart},%
  pdfauthor={Simon Pepping},%
  pdfsubject={The preprint document class elsart},%
  pdfkeywords={instructions for use, elsart, document class},%
  pdfstartview=FitH,%
  bookmarks=true,%
  bookmarksopen=true,%
  breaklinks=true,%
  colorlinks=true,%
  linkcolor=blue,anchorcolor=blue,%
  citecolor=blue,filecolor=blue,%
  menucolor=blue,pagecolor=blue,%
  urlcolor=blue]{hyperref}
\else
\usepackage[%
  breaklinks=true,%
  colorlinks=true,%
  linkcolor=blue,anchorcolor=blue,%
  citecolor=blue,filecolor=blue,%
  menucolor=blue,pagecolor=blue,%
  urlcolor=blue]{hyperref}
\fi

\newcommand{\ket}[1]{\left\vert#1\right\rangle}
\newcommand{\bra}[1]{\left\langle#1\right\vert}

\newcommand{\dmat}[2]{\ket{#1}\!\!\bra{#2}}
\newcommand{\tr}{\mbox{Tr}}

\newcommand{\nv}{\vec{n}}
\newcommand{\nsn}{\vec{n}\star\vec{n}}
\newcommand{\ndn}{\vec{n}\cdot\vec{n}}
\newcommand{\nsndn}{\vec{n}\star\vec{n}\cdot\vec{n}}

\newcommand{\beq}{\begin{equation}}
\newcommand{\eeq}{\end{equation}}
\newcommand{\bea}{\begin{eqnarray}}
\newcommand{\eea}{\end{eqnarray}}

\makeatletter
\def\elsartstyle{%
    \def\normalsize{\@setfontsize\normalsize\@xiipt{14.5}}
    \def\small{\@setfontsize\small\@xipt{13.6}}
    \let\footnotesize=\small
    \def\large{\@setfontsize\large\@xivpt{18}}
    \def\Large{\@setfontsize\Large\@xviipt{22}}
    \skip\@mpfootins = 18\p@ \@plus 2\p@
    \normalsize
}
\@ifundefined{square}{}{\let\Box\square}
\makeatother

\begin{document}

\begin{frontmatter}

\title{General Depolarized Pure States: Identification and Properties}

\author[Byrd]{Mark S. Byrd}
\ead{mbyrd@physics.siu.edu}
\author[Brennen]{Gavin K. Brennen}
\address[Byrd]{Department of Physics and Department of Computer Science, Southern Illinois University, Carbondale, Illinois 62901-4401}
\address[Brennen]{Institute for Quantum Optics and Quantum Information of the 
Austrian Academy of Sciences, A-6020, Innsbruck, Austria}

\begin{abstract}
The Schmidt decomposition is an important tool 
in the study of quantum systems especially for the quantification 
of the entanglement of pure states. However, the Schmidt decomposition 
is only unique for bipartite pure states, 
and some multipartite pure states. Here 
a generalized Schmidt decomposition is given for 
states which are equivalent to depolarized pure states.  
Experimental methods for the 
identification of this class of mixed states are provided and 
some examples are discussed which show the utility of this 
description.  A particularly interesting example provides, 
for the first time, an interpretation of the number of negative 
eigenvalues of the density matrix.  
\end{abstract}

\end{frontmatter}


\section{Introduction}

Describing and quantifying entangled quantum states and preventing 
errors during quantum dynamical processes are 
important and, at this time, unsolved problems.  
Each of these has important implications 
for the development of reliable quantum information processing 
devices.  In order to tackle these problems, it is common 
to broaden 
our knowledge and understanding by developing key examples.  
This is our approach here as we examine a particular class of 
quantum states.  

This work was motivated by a desire to be able to identify 
and distinguish a certain class of mixed quantum states, 
and their properties, experimentally.
This will rely, in part, on the existence of the
Schmidt decomposition \cite{Schmidt:06} which provides a 
canonical form for bipartite pure states.  
The Schmidt decomposition 
is used to identify and quantify entanglement in 
bipartite quantum systems \cite{Nielsen/Chuang:book}.  Such systems 
are primitives for a host of quantum communication and computation
protocols.  However, 
such protocols are invariably subject to noise which diminishes 
their advantage over classical protocols.  Noise, for 
most quantum systems, is difficult 
to identify and protect against, although there are several 
promising methods (see for e.g.
\cite{Byrd/etal:pqe04} and references therein). 
Here we introduce 
a generalized Schmidt decomposition for a class of mixed 
quantum states which we hope will 
aide both with the problem 
of understanding entanglement and our 
ability to correct for noisy quantum processes.
Our decomposition does not retain all of the 
useful properties which make the pure-state version so 
important.  However, it does allow us to devise some 
useful tools for measuring properties of an 
important class of states.  

The Schmidt decomposition is described by a 
set of real coefficients that is invariant under local unitary operations.
All entanglement measures on pure states, such as the von Neumann entropy 
of a reduced density operator, can be computed from this set.  
However, this decomposition is known only to exist for general 
bipartite pure states (see for example \cite{Peres:book}) 
and some multipartite pure states \cite{Thapliyal:99,Pati:00}.  
Therefore, 
quantifying entanglement in terms of this decomposition does not work 
in general. For mixed states, several entanglement
measures exist, most of which are difficult to calculate, though some 
interesting special cases for bipartite systems can be solved.  
For example, for two qubits one can calculate the 
Entanglement of Formation (EoF) \cite{Wootters:98} 
which is the amount of entanglement required to form a particular 
state. It is also known how to calculate the EoF for Werner states 
\cite{Vollbrecht/Werner:symm}, isotropic states \cite{Terhal/Voll:iso} 
and rotationally invariant states \cite{Manne:05}.  
However, at this time there is no canonical Schmidt decomposition 
for mixed states and no efficient method by which to 
analytically compute the entanglement of general mixed states.  

One might anticipate that a 
generalization of the Schmidt decomposition would 
aid in the description of entangled states.  One such 
generalization is given by the Schmidt number \cite{Terhal:00sn}, 
which is equal to the maximum Schmidt rank (or number of 
Schmidt coefficients) in a pure state decomposition of a 
mixed state, minimized 
over all decompositions.  This quantity constitutes the minimum 
Schmidt rank of the pure 
states needed to construct a state, and is an entanglement 
monotone \cite{Terhal:00sn}.  
Here we consider another special case which is a 
Schmidt decomposition for depolarized pure states (DPS) 
which are those states obtained by mixing 
the identity operator 
on the state space with a single pure state.  These have
many interesting properties and have been studied in the 
literature since these states are fairly easy to manipulate.
For example, one may compute properties such as 
channel capacities \cite{King:03,Datta:05}, 
entanglement (specific instances) 
\cite{Caves/Milburn:99,Rungta/etal:00}, and more recently, 
it has been shown that noisy operations may be turned 
into depolarizing operations 
\cite{Dur/etal:05}.  The set of DPS which we define here 
includes, not only 
pure states which have undergone a depolarizing operation, 
but also states which, 
if initially decoupled from their environment, cannot be 
obtained in this way.  All states in our DPS class 
can be brought into a similar canonical form using local 
unitary operations.  

The DPS are important to understand in part because 
they have a fairly simple form.  This form has $2D-1$ real parameters as
opposed to $D^2-1$ parameters for a generic mixed state in a $D$ 
dimensional Hilbert space.  They are also important to 
understand because 
any map can be brought to the depolarizing form by a simple 
sequence of quantum operations.  Therefore a 
complicated quantum computing process in the presence of noise 
can be brought into this form which produces states with 
relatively few relevant parameters.  This allows a direct 
comparison of inequivalent noise processes by projecting them 
into the same class.

In this article we discuss methods for experimentally determining 
whether this form has indeed been produced.  We find expressions 
for the fidelity and the trace distance for this class of mixed 
states, and are also able to show that the negativity is 
more easily quantified for bipartite DPS.  More 
importantly perhaps, {\it we provide a bound for the 
number of negative eigenvalues for bipartite DPS and 
show that the number of negative eigenvalues can 
indicate the type of entanglement present in the system}, 
e.g. qubit-qubit vs. qutrit-qutrit.  These results 
support a limited form of a conjecture by Han, et al. \cite{Han/etal:06} 
about the maximum number of negative eigenvalues for a bipartite state.  
We emphasize that our results provide an experimentally detectable 
qualitative and quantitative measure of entanglement.

The paper is organized as follows.  
In Section \ref{sec:cohvec} we review the coherence vector 
parameterization of the density operator.  In Section 
\ref{sec:dpc} we provide a geometric interpretation 
of DPS in terms of the coherence vector parameterization.  
Section \ref{sec:schmidtpf} demonstrates that there exists a type 
of Schmidt decomposition for depolarized pure states when 
there exists a Schmidt decomposition for the corresponding 
pure state.  In Section \ref{sec:expid} we provide two 
ways in which to identify these states experimentally, and 
describe physical maps which give rise to DPS beginning 
in an unknown pure state.  In Section 
\ref{sec:ent} we discuss the insight that we gain into 
bipartite entanglement given our construction.  We then conclude 
with a summary and some open questions in Section \ref{sec:concl}.  
Some examples of the formalism are given in Appendix A.


\section{Schmidt form for DPS}

In this section we provide several forms for the DPS which will 
be used for various calculations in later sections.  


\subsection{The coherence, or Bloch, vector}

\label{sec:cohvec}

The generalized coherence vector, or Bloch vector representation 
\cite{Mahler:book,Jakob:01,Byrd/Khaneja:03,Kimura} 
will provide a convenient geometric picture for several parts of 
our argument.  For a two-state system the description is well-known. 
The general case for an $D$-dimensional system is presented here 
and the two-state system will be seen to be a special case.  

Any density operator $\rho$ belonging to the set of bounded 
linear operators $\mathcal{B}(\rho)$ with Hilbert space dimension 
${\rm dim}(\mathcal{H})=D$, can be expanded in a basis 
consisting of the identity operator and an operator basis for 
$\mathfrak{su}(D)$, the algebra of $SU(D)$.  Throughout this 
work, we represent the latter with a set of Hermitian, traceless 
matrices, $\{\lambda_i\}_{i=1}^{D^2-1}$ which obey the following 
orthogonality condition
\beq
\mbox{Tr}(\lambda_i\lambda_j) = 2\delta_{ij}.
\eeq
The commutation and anticommutation relations for this set 
are summarized by the following product formula
\beq
\lambda_i \lambda_j = \frac{2}{D}\delta_{ij}{\bf 1}_D + i c_{ijk} \lambda_k 
                      + d_{ijk}\lambda_k.
\eeq
Here, ${\bf 1}_D$ is the $D\times D$ unit matrix, 
the $c_{ijk}$ are the structure constants of the Lie algebra 
represented by these matrices, and the $d_{ijk}$ are referred to 
as the components of the totally symmetric ``$d$-tensor.''  

The density matrix for an $D$-state system can now be written in 
the following form
\beq
\label{eq:ndmat}
\rho = \frac{1}{D}\left({\bf 1}_{D} 
          + \sqrt{\frac{D(D-1)}{2}} \; \vec{n}\cdot \vec{\lambda}\right),
\eeq
where $\vec{n}\cdot\vec{\lambda} = \sum_1^{D^2-1}n_i\lambda_i$. 
For $D>2$ the following conditions 
characterize the set of all pure states, 
\beq
\label{eq:pscond}
\vec{n}\cdot \vec{n} = 1, \;\;\; \mbox{and} \;\;\; 
          \vec{n}\star \vec{n} = \vec{n},
\eeq
where the ``star'' product is defined by
\beq
(\vec{a}\star \vec{b})_k = 
                   \sqrt{\frac{D(D-1)}{2}}\;\frac{1}{D-2} d_{ijk} a_i b_j.
\eeq
For $D=2$, the condition $\vec{n}\cdot \vec{n} = 1$
alone is sufficient \cite{Blochnote}.  Note that 
\beq
\label{eq:ns}
n_i = \sqrt{\frac{D}{2(D-1)}}\;\;\tr\left(\rho\lambda_i\right).
\eeq
To recover the case of the two-state Bloch sphere, note that the 
constants $1/D$ and $\sqrt{D(D-1)/2}$ reduce to $1/2$ and $1$ 
respectively, and the $d_{ijk}$ are identically zero, so the second 
condition in Eq.(\ref{eq:pscond}) is not required.  In fact, as 
noted, it cannot be satisfied.


\subsection{Depolarized Pure States}

\label{sec:dpc}

Throughout this paper we focus on a special class of 
mixed states, the depolarized pure states (DPS).  Such states are 
given by a (not necessarily convex) sum of the 
identity operator and a pure state:
\beq
\rho_d \equiv (1-p)\frac{1}{D}{\bf 1}_D + p\rho',
\label{eq:depmap}
\eeq
for $\rho'$ some pure state.  By the unit trace and positivity 
conditions, we have $-1/(D-1)\leq p\leq 1$.
Letting $c_D = \sqrt{D(D-1)/2}$, we may rewrite this in a more 
suggestive form as 
\beq
\label{eq:cvdep}
\rho_d = \frac{1}{D}\left({\bf 1}_D 
              + c_D \;p \vec{n}\cdot \vec{\lambda}\right). 
\eeq
We note that for $D>2$ the characterization is unique, i.e. $\rho_d$
corresponds to a depolarized form of a single pure state with coherence 
vector $\vec{n}$.  This is because the condition $\vec{n}\star \vec{n}=\vec{n}$
demands that both $\vec{n}$ and $-\vec{n}$ cannot correspond to physical pure 
states.  Hence, any vector of the form $p\vec{n}$ has a unique 
purification, namely $\vec{n}$.  For $D=2$ this is not the case 
because both $\vec{n}$ and $-\vec{n}$ correspond to pure states.  
From this latter form, we may interpret the DPS as arising from 
the affine map:  $\vec{n}\mapsto p\vec{n}$, on the $D^2-1$ dimensional 
real vector space of coherence vectors.  

This provides a geometric description of the set of depolarized pure 
states.  
{\it The space of DPS with a given 
$p$ is isomorphic to the set of pure states (for $D>2$). } 
(See for example \cite{us} and references therein.)    
To see the geometry more 
explicitly, note that the DPS can be written in the form
\[
\label{eq:pdiag}
\rho_d = \frac{1}{D}\left({\bf 1}_D - pW\left[\begin{array}{ccccc}
                                              1 & & & & \\
					        &1& & & \\
					        & &\ddots& & \\
					        & &     & 1 & \\
					        & &     & & -(D-1)
					   \end{array}\right]W^\dagger\right).
\]
Note that the same matrix $W$ will diagonalize both the pure state 
and the depolarized pure state.   


We will make use of this form to 
{\it analytically} compute the trace distance and 
fidelity between two DPS.  The fidelity between two density matrices is 
defined by 
\beq
F(\rho,\sigma)=\tr \Big[\sqrt{\sqrt{\rho}\sigma\sqrt{\rho}}\Big]^2.
\eeq
We consider two DPS both in a $D$ dimensional Hilbert space,
\[
\begin{array}{lll}
\rho_d&=&(1-p)\frac{{\bf 1}}{D}+p\ket{\Psi}\bra{\Psi}\\
\sigma_d&=&(1-q)\frac{{\bf 1}}{D}+q\ket{\Phi}\bra{\Phi}\\
\end{array}
\]
where $-\frac{1}{D-1}\leq p \leq 1$ and the overlap in their purifications is
$F(\ket{\Psi},\ket{\Phi})=|\langle\Psi\ket{\Phi}|^2=f$.  The 
(square root) of the fidelity is
\beq
\begin{array}{lll}
\sqrt{F(\rho_d,\sigma_d)}&=&(D-2)\sqrt{a}+\sum_{\pm}\Bigg[\frac{2 a+(b+2c)f+d+b(1-f)}{2}\pm\\
&&\sqrt{\frac{((b+2c)f+d-b(1-f))^2}{4}+(b+c)^2(1-f)f}\Bigg]^{\frac{1}{2}},\\
\end{array}
\eeq
where the parameters are given by:
\[
\begin{array}{lll}
a&=&\frac{(1-p)(1-q)}{D^2},\\
b&=&\frac{(1-p)q}{D},\\
c&=&\frac{q}{D}\big(\sqrt{((D-1)p+1)(1-p)}-(1-p)\big),\\
d&=&\frac{(1-q+Dqf)}{D^2}\big((D-2)p+2-2\sqrt{((D-1)p+1)(1-p)}\big)\\
&&+\frac{2(1-q)}{D^2}\big(\sqrt{((D-1)p+1)(1-p)}-(1-p)\big).
\end{array}
\]
\begin{figure}
\begin{center}
\includegraphics[scale=0.5]{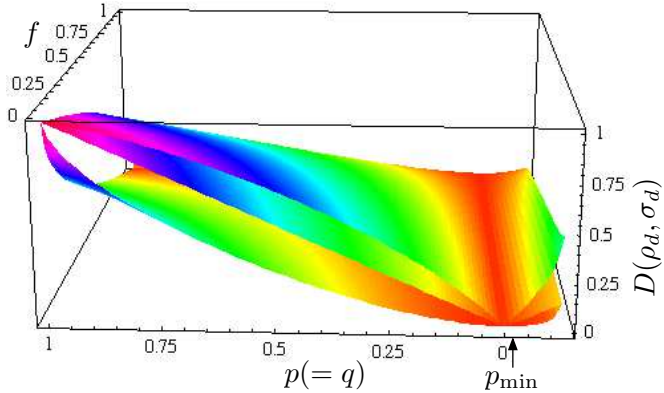}
\caption{Distance measures between two depolarized pure states (DPS) in a $D=9$ dimensional Hilbert space with equal polarizations $p=q$.  The Bures metric, trace distance, and fidelity satisfy the following inequalities \cite{Fuchs:99}:  $B(\rho,\sigma)^2/2\leq D(\rho,\sigma)\leq \sqrt{1-F(\rho,\sigma)}$ and surface plots of these three functions are shown.  The value $p_{\rm min}=-1/80$ is the minimum polarization of depolarized pure states which are obtainable from a 
completely positive map. (We call these physically depolarized pure states, cf. Sec.~\ref{sec:Physdep}.)}
\label{fig:1}
\end{center}
\end{figure}
The square root of the fidelity can be converted into a metric, 
specifically the \emph{Bures metric} via 
$B(\rho,\sigma)=\sqrt{2-2\sqrt{F(\rho,\sigma)}}$, 
and an angle $A(\rho,\sigma)=\cos^{-1}\sqrt{F(\rho,\sigma)}$.
In the pure state case, the Bures metric is the Euclidean distance 
between the two pure states with respect to the 
norm on the state space and the cosine of the angle between the 
states is the overlap.  The Bures metric between 
two mixed states can be 
interpreted as the Euclidean distance between purifications
of the mixed states minimized over all such purifications.  

One can also compute the distance (in the trace norm) between two 
mixed states.  The distance is
\beq
D(\rho,\sigma)=\frac{1}{2}\tr |\rho-\sigma|_{\rm tr},
\eeq
where the trace norm is defined $|O|_{\rm tr}=\sqrt{O^{\dagger}O}$.   
For the two DPS,
\beq
\begin{array}{lll}
D(\rho_d,\sigma_d)&=&\frac{1}{2}\Bigg[ \frac{(D-2)|q-p|}{D}+\sum_{\pm}\Big| \frac{(q-p)(1-D/2)}{D}\\
&&\pm\sqrt{(\frac{p+q-2 q f}{2})^2+q^2(1-f)f}\Big|\Bigg]\\
\end{array}
\eeq
The distance between two mixed states with the same coherence 
vector magnitude $p=q$ is simply
$D(\rho_d,\sigma_d)=(1-f)|p|$.  The distance and fidelities 
of equally polarized pure states are plotted 
in Fig. \ref{fig:1}.  Notice that beginning in a pure state, 
i.e. $p=1$, the distance and Bures metric between states
with $f<1$ will decrease under a depolarizing map until both 
states are mapped to the identity.
For even stronger maps, i.e. $p<0$ the distance begins to increase 
again.  As discussed in Sec.~\ref{sec:Physdep},
the minimum value of polarization obtainable by a physical map 
acting on input pure states is $p_{\rm min}=-1/(D^2-1)$.
At this value, the distance between the output states is 
$D(\rho_d,\sigma_d)=\frac{1-f}{D^2-1}$.  Thus we find 
that the distance (fidelity) between initially pure states is not a 
monotonically decreasing (increasing) function of the 
depolarization strength.






\subsection{Schmidt Decomposition For A Pure Bipartite State}

\label{sec:schmidtpf}

To fix notation, 
let us recall the Schmidt decomposition for 
a pure state of a bipartite quantum 
system in $D = D_A \times D_B$ dimensions with subsystems 
$A$ and $B$ which have dimension $D_A$ and $D_B$ respectively.  
Without loss of generality, we will assume that $D_A\leq D_B$.  
Now, let 
\beq
\rho_{AB}^{\prime} = \dmat{\Psi^\prime}{\Psi^\prime},
\eeq
where 
\beq
\ket{\Psi^\prime} = \sum_{i,\mu} a_{i\mu}\ket{{\phi_i}^\prime}_A\ket{{\psi_\mu}^\prime}_B.
\eeq
According to the Schmidt decomposition \cite{Schmidt:06}, there 
exist unitary matrices $U$ which acts only on the first subsystem, 
and $V$ which acts only on the second 
subsystem, such that $\ket{\Psi}$ can be written in the form:
\beq
\label{eq:Schmidtform}
\ket{\Psi} = \sum_{j} b_{j}\ket{\phi_j}_A\ket{\psi_j}_B,
\eeq
where the set $\{\ket{\phi}_A\}$ ( $\{\ket{\psi}_B\}$) forms an 
orthonormal basis for $A$ ($B$). 
In other words, there are local unitary transformations, 
$U$ and $V$ such that 
\beq
U\otimes V\ket{\Psi^\prime} = \sum_{j} b_{j}\ket{\phi_j}_A\ket{\psi_j}_B,
\eeq
where 
\beq
b_{j}\delta_{j\nu} 
= \sum_{i,\mu}U_{ij}a_{i\mu}V_{\mu\nu}.
\eeq
and $U,V$ can chosen so that the $\{b_j\}$ are real
and positive.
We will say that $a_{i\mu}$ is ``diagonalized''  \cite{diagnote}
by the local unitary transformations $U$ and $V$. 
The reduced density matrices 
$\rho_A = \tr_B (\rho) = \sum_j b_j^2 \ket{\phi_j}_A\bra{\phi_j}$ 
and 
$\rho_B = \tr_A (\rho)= \sum_j b_j ^2 \ket{\psi_j}_B\bra{\psi_j}$ 
have the same eigenvalues $b_j^2$.  




Now, let us consider the density operator 
\beq
\rho^{\prime} = \frac{{\bf 1}_{AB}}{D} +  \left(%
             \sum_{ik,\mu\beta} a_{i\mu}a_{k\beta}^*
               \ket{\phi_i^\prime}_A\bra{\phi_k^\prime}\otimes %
    \ket{\psi_\mu^\prime}_B\bra{\psi_\beta^\prime}-\frac{{\bf 1}_{AB}}{D}\right).
\eeq
Defining the matrix $\sqrt{|p|}\;a_{i\mu}\equiv c_{i\mu}$, we see that 
if the matrix $a_{i\mu}$ can be diagonalized by $U$ and $V$, 
then $\sqrt{|p|} \;a_{i\mu}$ can be diagonalized by 
the {\it same} $U$ and $V$. 
\beq
\rho_d = \frac{{\bf 1}_{AB}}{D}(1-p) +
      p\sum_{j,j'} b_jb_{j'} \dmat{\phi_j}{\phi_{j'}}\otimes\dmat{\psi_j}{\psi_{j'}}.
\eeq
Therefore, there exists a preferred 
local unitary basis for depolarized pure states and 
we refer to this preferred basis as the Schmidt decomposition 
for DPS.

Furthermore, we can provide a relationship between the 
eigenvalues of the reduced density matrices for the two subsystems.  
Tracing over the subsystem $B$ produces
\beq
\rho_{d_A} = \tr_B(\rho_d) 
           = \frac{{\bf 1}_A}{D_A}(1-p)+p\sum_j b_j^2\dmat{\phi_j}{\phi_j}.
\eeq
Now, let us suppose that there are $n$ non-zero eigenvalues of 
$\rho^{\prime}_A=\tr _B(\rho')$ given by $\{b_j^2\}$ with 
$\sum_{i=1}^n b_i^2 = 1$.  
(Alternatively, we could let the sum go to $D_A$ noting that for 
some $i$, the eigenvalue could be zero.)
Then the eigenvalues of $\rho^\prime_A$ are 
$\{\frac{1}{D_A}(1-p) +pb_i^2\}_{i=0}^{n-1}\sqcup\{(1-p)/D_A\}^{D_A-n}$. 
Tracing over the subsystem $A$ produces
\beq
\rho_{d_B} = \tr_A(\rho_d) 
           = \frac{{\bf 1}_B}{D_B}(1-p)+p\sum_j b_j^2\dmat{\psi_j}{\psi_j}.
\eeq
The eigenvalues of $\rho^\prime_B$ are given by 
$\{\frac{1}{D_B}(1-p) +pb_i^2\}_{i=0}^{n-1}\sqcup\{(1-p)/D_B\}^{D_B-n}$.

There are two properties of the Schmidt decomposition 
which make it particularly useful and are properties which one would 
want to preserve in any generalization.  It specifies (i) preferred 
bases of (ii) bi-orthogonal states.    
It is clear that property (i) is retained for DPS.  
This relies on the fact that it is unique 
for pure states \cite{Peres:book} 
barring a degeneracy in the spectrum of one of the subsystems.  

The Schmidt decomposition for general bipartite 
DPS is the preferred basis which agrees with the pure state 
Schmidt decomposition counterpart of the DPS. This 
definition clearly retains the property (i) and it 
can be generalized to any system with a corresponding 
pure state Schmidt decomposition.  For example those described 
by a multipartite Schmidt decomposition 
\cite{Thapliyal:99,Carteret:00} will also have corresponding 
set of DPS.  

Can this preferred basis be used to quantify the entanglement 
of the system?  Certainly this is not true for 
the entropy of the partial trace as can be seen by considering 
the extreme case where $p=0$.  However, we will discuss 
how the Schmidt form helps identify and distinguish 
certain types of entangled states in Section \ref{sec:ent}.


\section{Preparation and Identification of DPS}

\label{sec:expid}

It is now pertinent to ask, how does one know if a density matrix 
describes a system whose state is in the class DPS?
Is there a way to characterize maps which give rise 
to these states?  This section will provide the answers to 
these questions.  


\subsection{State Tomography}

Using state tomography the elements of the density matrix may be 
determined.  There are several ways in which to do this, some 
of which are more efficient than others.  
For our purposes, it is assumed that 
state tomography data has been collected and from it the 
coherence vector $\vec{n}$ determined, for example via 
Eq.~(\ref{eq:ns}).  

From Eq.(\ref{eq:pscond}) the coherence vector of a pure state 
satisfies $\nsn = \nv$.  For a DPS, $\nv\rightarrow p \nv$, so 
that $\nsn \rightarrow p^2\nsn$, etc.  From these relations, it 
is clear that all invariants described in 
\cite{Byrd/Khaneja:03} can be calculated by noting that 
for a DPS $[\nv\star]^r\nv\cdot\nv =p^{r+2}$. Therefore the 
invariants reduce to the simplified form which 
is obtained by replacing $\nv$ with $p$ everywhere and neglecting 
the types of products. In other words, 
\[
\begin{array}{lll}
\ndn &=& p^2,\\ 
\nsn\cdot\nv &=& p^3,\\
\nsn\star\nv\cdot\nv &=& p^4,\mbox{ and so on.}
\end{array}
\]
These conditions may be stated equivalently, and more 
succinctly, as 
\beq
\label{eq:depolconditions}
\ndn = p^2, \;\;\; \mbox{and} \;\;\; \nsn = p\nv. 
\eeq
Note that, similar to the pure state conditions, these two 
conditions alone determine the set of 
eigenvalues for the density operator.  

Note also that the DPS with $p<0$ and with $p>0$ 
can be distinguished with the unitary invariant 
$\nsndn$ (provided $D>2$).  Hence given some prior certificate
that the state is a DPS, we obtain complete spectral information
from the measurement of $\ndn$ and $\nsndn$ 
including the value of $p$.  

Alternatively, one may examine the eigenvalues of the system. 
If the eigenvalues are given by 
$a,b,b,...,b$ and having $a+(D-1)b = 1$, 
then the system is in the class DPS.  Notice that the spectrum of 
the bipartite density matrix can be used to define 
the class and this is unchanged by a global unitary 
transformation.


\subsection{Invariant Polynomials}

Another measurement process which will efficiently 
identify the DPS is due to Brun \cite{Brun:04}. 
He showed that, in principle, 
the invariants $\tr(\rho^m)$ could be measured efficiently. 
From these, the eigenvalues may be determined.  

Let $\hat{S}$ be an operator which cyclicly permutes states of the 
system:
\beq
\hat{S}\ket{\psi_1}\ket{\psi_2}\cdots\ket{\psi_n} 
          = \ket{\psi_n}\ket{\psi_1}\ket{\psi_2}\cdots\ket{\psi_{n-1}}, 
\eeq
then 
\beq
\tr(\hat{S}\rho^{\otimes m}) = \tr(\rho^m).
\eeq
To show this is quite straight-forward.  Let 
\beq
\rho=\sum_i p_i \rho_i = \sum_i p_i \dmat{\psi^i}{\psi^i}
\eeq
be an orthogonal ($\tr(\rho^i\rho^j)=\delta^{ij}$) 
pure-state decomposition of the density matrix. Then 
\beq
\begin{array}{lll}
\hat{S}\rho^{\otimes m} 
     &=& \rho_{1}\otimes\rho_{2}\otimes\cdots \rho_{m}\nonumber \\
     &=& \hat{S}\sum_{i_1}\sum_{i_2} \cdots \sum_{i_m} p_{i_1}p_{i_2} ...p_{i_m}
        \dmat{\psi^{i_1}_1}{\psi^{i_1}_1}\otimes\dmat{\psi^{i_2}_2}{\psi^{i_2}_2}\\
        &&\otimes\cdots\otimes \dmat{\psi^{i_m}_m}{\psi^{i_m}_m} \\
     &=& \sum_{i_1}\sum_{i_2} \cdots \sum_{i_m} p_{i_1}p_{i_2} ...p_{i_m}
        \ket{\psi^{i_m}_m}\ket{\psi^{i_1}_1}\ket{\psi^{i_2}_2}\cdots
   \ket{\psi^{i_{m-1}}_{m-1}} \\
   &&\bra{\psi^{i_1}_1}\bra{\psi^{i_2}_2}
                 \cdots\bra{\psi^{i_m}_{m}}.
\end{array}
\eeq
Taking the trace simply produces a series of Kronecker deltas which 
force all $p_i$ to have the same index so that 
\beq
\tr(\hat{S}\rho^{\otimes m}) = \sum_ip_i^m = \tr(\rho^m). 
\eeq
A physical implementation of this measurement can be realized 
using an interferometer type circuit.  This works by preparing an ancilla 
qubit $a$ in the state 
$|+_x\rangle_a, (|\pm_x\rangle_a=1/\sqrt{2}(|0\rangle_a\pm|1\rangle_a))$,
and applying 
a sequence of $m-1$ controlled-SWAP gates between the ancilla and 
pairs of copies of $\rho$:
\[U=\prod_{j=0}^{m-2}\ket{0}_a\bra{0}\otimes {\bf 1}_{1\ldots m}+\ket{1}_a\bra{1}\otimes 
\mbox{SWAP}(m-j,m-j-1),
\]
where $\mbox{SWAP}(r,s)=\sum_{i,j=0}^{D^2-1}\ket{i}_r\bra{j}\otimes \ket{j}_s\bra{i}$.
Each controlled-SWAP gate can be implemented using $O(D^2)$
elementary two qudit gates \cite{BBO:05}. 
A final measurement of the ancilla in the $\ket{\pm _x}_a$ basis
gives measurement outcomes $m=\pm 1$ with probability 
$P(m=\pm 1)=\frac{1}{2}(1+\tr[\rho^m])$.    

Since the above result really only depends on the production of the 
appropriate delta functions, in practice, 
any cyclic permutation which is not 
the identity could be used. In fact, it need not be cyclic 
as long as there is no invariant subspace.

One may suppose that a particular experiment may provide for 
a more efficient measurement using the polynomials.  However, 
it may also be the case that some state tomography data is 
available or some partial information about the state is known,  
In either of these cases, it is relevant to note the 
$\tr(\rho^m)$ and the coherence/Bloch vector are directly 
related \cite{Byrd/Khaneja:03,Kimura}.


\subsection{Efficient determination using local measurements}

Knowing that a system is in a DPS enables the determination of 
the eigenvalues of $\rho_d$ with the determination of $\tr(\rho_d^2)$ 
and $\tr(\rho_d^3)$ alone.  However, if we do not know whether or 
not the combined system is in a DPS, a natural question is, 
how could this be determined?  Generically this could be achieved 
by measuring the full spectrum of the state as outlined above by 
performing $D$ measurements 
over a total of $D(D+1)/2$ identically prepared copies of 
the state.  For bipartite systems, simpler measurements on the subsystems
$A$ and $B$ can reveal partial information about the state.  While 
such information is not sufficient to verify that the joint state is of 
DPS form, one can check for a violation of the consistency relations 
given in Sec.~\ref{sec:schmidtpf} 
that can rule out that possibility.  
For example, one can measure the spectrum of the reduced 
states $\rho_A,\rho_B$ and verify that the two sets of eigenvalues
are equal up to the scaling which depends on the dimension.
Another, perhaps simpler, measurement is to verify that the density 
operators are full rank.  If one reduced state was found to 
have rank less than its dimension, 
for example by obtaining a zero value in a projective
measurement, then the corresponding 
combined state $\rho_{AB}$ could not be a DPS.  
Furthermore, for $D_B \geq D_A+2$, there must exist a degenerate 
subspace of the subsystem $B$ of dimension $D_B-D_A$.  If this 
is not present, the system cannot be in a DPS.


\subsection{Physical depolarization channels}
\label{sec:Physdep}

It is natural to ask if all states $\rho_d$ 
can be generated 
by beginning in a pure state $\rho'$ and applying a physical map
which depolarizes that state to the form $\rho_d$.
It turns out that this is not 
always possible.  Rather, according to the value of $p$, there is a 
continuous subset of DPS that cannot be so generated.
To see this, consider the class of maps
\beq
\label{eq:gdepmap}
{\cal E}_p(\rho) = (1-p)\frac{1}{D}{\bf 1}_D + p\rho\equiv \rho.
\eeq
In ref.~\cite{Rungta:01a} it was shown that 
maps $\mathcal{E}_p$ with $ -1/(D-1)\leq p \leq 1$ are positive,
but only those with $ -1/(D^2-1)\leq p \leq 1$  are completely
positive.  Completely positive maps (CPM) are those maps which 
act as the identity operator on an environment when the input 
is a tensor product state of the system and environment.  Such maps are 
deemed to be physically allowed maps acting on a system which is
uncorrelated with its environment. (However, some dynamics 
need not be completely positive 
\cite{Pechukas:94,Pechukas+Alicki:95,Jordan:04}.) The map 
$\mathcal{E}_{p=-1/(D^2-1)}$ 
is termed the universal inverter as it outputs the positive 
operator closest to being an inversion of the coherence vector of 
an arbitrary input state.
Given this demarcation 
we classify all states $\rho_d$ which are obtainable from a single 
copy of the (generically unknown) pure state $\rho'$ via a CPM to 
be physically depolarized pure states (PDPS).  
The criterion that the map act only on a single copy is emphasized because
more powerful operations are possible using multiple copies.  For example, 
given an infinite number of copies of a pure state $\rho'$ one CPM is 
to perform state tomography and from the classical information, 
synthesize $\rho_d$ exactly.  

One can synthesize any positive density operator $\rho_S$ in a $D$ dimensional 
Hilbert space by preparing an entangled state of the system with a $D$ dimensional 
ancilla $a$ and tracing over the ancilla.  Namely, given an eigen-decomposition of 
the state 
$\rho_S=\sum_j p_j \ket{\psi_j}_S\bra{\psi_j}$, one prepares the pure state 
$\ket{\Psi}_{Sa}=\sum_j \sqrt{p_j}\ket{\psi_j}_S\ket{j}_{a}$, and traces 
over the ancilla.  Clearly this synthesizes 
any DPS.  Yet, for an initially uncorrelated 
system and environment, the 
transformation is generically non-linear.  
Often it is the case that one is interested in generating 
a PDPS output given an unknown pure
state $\rho'_S$ as input.  This can be useful to drive noisy maps with 
many parameters on pure states, to a standard form of a quantum channel 
with only one parameter, namely $p$.  We now discuss two protocols to do so.  

The first method is a variant of a construction in \cite{Rungta:01a}.  Here 
one performs joint operations on the system and two ancillary qudits $a_1$ and $a_2$ each 
of dimension $D$.  The initial state is a tensor product state 
of the system $S$ and the ancillae:
\beq
\rho=\rho'_{S}\otimes \ket{\chi}_{a_1 a_2}\bra{\chi}
\eeq
where 
$\ket{\chi}_{a_1 a_2}=\alpha \ket{\Phi^+}_{a_1 a_2}+\beta 
\ket{0}_{a_1}\frac{1}{\sqrt{D}}\sum_{j=0}^{D-1} \ket{j}_{a_2}$, and 
$\ket{\Phi^+}=\frac{1}{\sqrt{D}}\sum_{j=0}^{D-1}\ket{j}\ket{j}$ is the 
maximally entangled state.  The parameter $\alpha$ can arbitrarily be chosen real.
We are interested in the case where the system itself is composed of two 
parts $A$ and $B$ but for simplicity we treat it as a single system whose 
Hilbert space is
spanned by the orthonormal states $\{\ket{j}_S\}_{j=0}^{D-1}$.
The next step is to apply a unitary composed of pairwise coupling gates 
between qudits:
\beq
\begin{array}{lll}
U_{Sa_1a_2}&=&[\prod_{j}^{D-1}X^j_S \otimes \ket{j}_{a_2}\bra{j}]
[\prod_{j'}^{D-1}X^{\dagger\ j'}_S \otimes \ket{j'}_{a_1}\bra{j'}]\\
&&
[\prod_{j}^{D-1}\ket{j}_{S}\bra{j}\otimes Z^{\dagger\ j}_{a_1} ]
[\prod_{j'}^{D-1}\ket{j}_{S}\bra{j}\otimes Z^{j'}_{a_2} ].\\
\end{array}
\eeq
Here the unitary operators are defined $X=\sum_j \ket{j+1}\bra{j}$ 
and $Z=\sum_j e^{i2\pi j/D}\ket{j}\bra{j}$.
The action of this unitary on a pure state input for the system is 
$U_{Sa_1a_2}\ket{\psi}_S\ket{\chi}_{a_1a_2}=\alpha \ket{\psi}_{A_1}\ket{\Phi^+}_{a_1a_2}+\beta 
\ket{\psi}_{a_1}\ket{\Phi^+}_{Sa_2}$.  Upon tracing over the ancillae, the residual system
state is then:
\beq
\begin{array}{lll}
\rho_{S}&=&\tr_{a_1a_2}[U_{Sa_1a_2}\rho U^{\dagger}_{Sa_1a_2}]\\
&=&(1-|\beta|^2)\rho'_S+|\beta|^2\frac{{\bf 1}_S}{D},
\end{array}
\eeq
where by the normalization constraint on the state $\ket{\chi}$, $0\leq |\beta|^2\leq D^2/(D^2-1)$.
Hence, by varying the parameter $\beta$, one can realize any PDPS.

A second protocol for generating PDPS works 
by using stochastic unitaries to randomize a quantum operation
$\mathcal{E}$ on an input state \cite{Dur/etal:05}.  
The degree to which the map $\mathcal{E}$
acts trivially determines the depolarization parameter $p$ and the randomization guarantees 
that the map takes all inputs $\rho$ to the standard form $\rho_p$.  
Specifically, one randomly picks a unitary $U\in U(D)$ and applies $U$ before and $U^{\dagger}$ after a trace preserving, CPM $\mathcal{E}$ on the state.  The result is
 \beq
 \begin{array}{lll}
 \mathcal{E}'(\rho)&=&\int dU U\mathcal{E}(U^{\dagger}\rho U)U^{\dagger}\\
 &=&\frac{D^2f-1}{D^2-1}\rho+\frac{D^2(1-f)}{D^2-1}\frac{{\bf 1}}{D}
 \end{array}
 \eeq
where $dU$ is the invariant Haar measure on $U(D)$.  Here $0\leq f\leq 1$ 
quantifies the identity portion of the map, i.e.  
$f=\bra{\Phi^+}E_{\mathcal{E}}\ket{\Phi^+}$ where $E_{\mathcal{E}}$
is the Choi-Jamio\l kowski representation \cite{Choi:75,Jamiolkowski:72} 
of the map $\mathcal{E}$.  Such a representation arises by first writing 
a trace preserving CPM on $\mathcal{B}(\mathcal{H_S})$ in a particular 
operator-sum decomposition as 
$\mathcal{E}(\rho)=\sum_{m,n,m',n'=0}^{D-1}E_{m,n;m',n'}X^nZ^m\rho (X^{n'}Z^{m'})^{\dagger}$.  
The state $E\in\mathcal{B}(\mathcal{H}_S\times \mathcal{H}_{S'})$ 
given by $E=\sum_{m,n,m',n'=0}^{D-1}E_{mn,m'n'}\ket{\Phi_{m,n}}_{SS'}\bra{\Phi_{m',n'}}$ expanded in the orthonormal basis $\{\ket{\Phi_{m,n}}_{SS'}=X_S^nZ_S^m\ket{\Phi^{+}}_{SS'}\}$,
is then the Jamio\l kowski representation of $\mathcal{E}$.  This follows by virtue of the 
relation $E_{\mathcal{E}}= \mathcal{E}_{S} \otimes {\bf 1}_{S'}(\ket{\Phi^+}_{SS'}\bra{\Phi^+})$.

A simple way to generate a particular PDPS is as follows:
\begin{itemize}
\item  Begin with a pure state $\rho'$.
\item  Pick a unitary $U\in U(d)$ at random and apply it to the state.
\item  Apply a quantum operation with Jamio\l kowski fidelity $f$;  for 
example, the single qudit unitary $V=e^{i\alpha (X_A+X^{\dagger}_A)}$ 
which has
$f=\frac{1}{D_A^2}|\sum_{j=0}^{D_A-1}e^{i 2\alpha\cos(2\pi j/D_A)}|^2$.
Another option is to apply the operator $X_A$ with probability
$1-f$ and with probability $f$ do nothing to the state.
\item Apply $U^{\dagger}$ to the state. 
\end{itemize}
The resultant state is $\rho_d$ with $p=\frac{D^2f-1}{D^2-1}$.
In practice, for the stochastic process, it is not necessary to pick a 
unitary uniformly at random, rather one can pick a random unitary from 
the finite set $\tilde{G}=G\setminus{\bf 1}$, where $G$ is the Clifford 
group.  The latter is defined as the group which leaves the Pauli group 
$P=\{e^{i2\pi k/D}X^aZ^b; a,b,k\in \mathbb{Z}_D\}$ invariant under 
conjugation.

We stress that both of the above protocols require performing entangling 
operations between the subsystems $A$ and $B$. This is because in both 
cases, it is necessary to implement the Pauli operators $X_S$ and $Z_S$ 
which cannot be written as local unitaries on $A$ and $B$ alone.  This 
emphasizes the fact that the depolarizing map is a map on the joint space, 
it cannot be realized by separately depolarizing each party.  In fact the 
action of individual depolarization is a map with $4$ real parameters: 
\[
\alpha_{00}\rho'_{AB}+\alpha_{01}\rho'_A\otimes\frac{{\bf 1}}{D_B}+
\alpha_{10} \frac{{\bf 1}}{D_A}\otimes \rho'_B+\alpha_{11}\frac{{\bf 1}}{D}
\]
which is not the desired form.


\section{Entanglement of DPS }

\label{sec:ent}

Given the results of Section \ref{sec:expid}, we can determine 
experimentally whether the state has the form of a DPS or not.  
From this information we find the negative 
eigenvalues which provides a sufficient condition 
for the existence of entanglement in a mixed state.  For a two 
qubit system, or a qubit-qutrit system 
the criterion is both sufficient and necessary.   


\subsection{Partial Transpose}

Since partial transpose is independent of local unitary operations, 
we can compute it for the Schmidt form of a depolarized state.
The explicit form of the partially transposed state is:
\beq
\label{eq:rhopt}
\begin{array}{lll}
\rho_d^{T_B}&=&(1-p)\frac{{\bf 1}_{AB}}{D}+p\sum_{j,j'=0}^{D_A-1}b_jb_{j'}
\ket{\phi_j}_A\bra{\phi_{j'}}\otimes\ket{\psi_{j'}}_B\bra{\psi_j}\\
&=&(1-p)\frac{{\bf 1}_{AB}}{D}+p\sum_{j=0}^{D_A-1}b_j^2
\ket{\phi_j}_A\bra{\phi_{j}}\otimes\ket{\psi_{j}}_B\bra{\psi_j}\\
&&+p\displaystyle{\sum_{j<j'=0}^{D_A-1}}b_jb_{j'}(\ket{+_{j,j'}}_{AB}\bra{+_{j,j'}}-\ket{-_{j,j'}}_{AB}\bra{-_{j,j'}})
\end{array}
\eeq
where we introduced the orthonormal states:  
$\ket{\pm_{j,j'}}=(\ket{\phi_j}\ket{\psi_{j'}}%
                   \pm\ket{\phi_{j'}}\ket{\psi_{j}})/\sqrt{2}$.
Notice that this form is diagonal.  


\subsection{Negativity}

For states $\rho\in\mathcal{B}(\mathcal{H}_A\times 
\mathcal{H}_B)$ with $D_{A(B)}={\rm dim}\mathcal{H}_{A(B)}$ the negativity 
$\mathcal{N}(\rho)$ is defined \cite{Vidal:02}:
\beq
\mathcal{N}(\rho)=\frac{|\rho^{T_B}|_{\rm tr}-1}{D_A-1}
\eeq
where, again, without loss of generality we assume $D_A\leq D_B$.  
The function is real valued and 
normalized to lie in the range $[0,1]$.  The argument $\rho^{T_B}$ 
is the partial transpose of $\rho$ with respect to subsystem $B$, 
which in a coordinate representation with 
$\rho=\sum_{i,i',j,j'}\rho_{ij,i'j'}\ket{i}_A{_A}\bra{i'}\otimes \ket{j}_B{_B}\bra{j'}$, 
is $\rho^{T_B}=\sum_{i,i',j,j'}\rho_{ij,i'j'}\ket{i}_A{_A}\bra{i'}\otimes \ket{j'}_B{_B}\bra{j}$.
While it's action is locally basis dependent, the eigenvalues of 
$\rho^{T_B}$ are not, and the negativity counts a normalized
sum of the norm of negative eigenvalues.  Because any separable 
state can be written as a convex sum of products of 
partial density operators, 
and hence has eigenvalues invariant under partial transposition, 
negative eigenvalues are a sufficient 
\emph{but not necessary} condition for the presence of 
bipartite entanglement in $\rho$.
States with $\rho^{T_B}>0$ but not separable 
are known as bound entangled 
states because that entanglement cannot be distilled.

From Eq.~\ref{eq:rhopt} the negativity is quickly found to be:
\beq
\begin{array}{lll}
\mathcal{N}(\rho_d)&=&\frac{1}{D_A-1}\Big[(1-p)(1-\frac{D_A}{D_B})+\sum_{j=0}^{D_A-1} |\frac{1-p}{D}+pb_j^2|\\
&&+\sum_{j<j'=0}^{D_A-1}(|\frac{1-p}{D}+pb_jb_{j'}|+|\frac{1-p}{D}-pb_jb_{j'}|)-1\Big]\\
&=&\frac{1}{D_A-1}\Big[\sum_{j<j'=0}^{D_A-1}(pb_jb_{j'}+|\frac{1-p}{D}-pb_jb_{j'}|)\Big]-\frac{1-p}{2D_B}.\\
\end{array}
\eeq
All that is required for $\mathcal{N}(\rho_d)>0$ is that one of the 
terms inside the absolute value be negative or $p>\frac{1}{Db_jb_{j'}+1}$ 
for some pair of Schmidt coefficients $b_j,b_{j'}$.
Notice, that since $b_jb_{j'}\leq 1/2$, then for 
$p\leq \frac{1}{D/2+1}$, $\mathcal{N}(\rho_d)=0$.  It is also 
true that for $p\leq \frac{1}{D/2+1}$, 
the state is separable \cite{Horodecki:00}.  

However, let us note that, from the diagonal form, we can 
extract more information.  Any quantifier of entanglement, such 
as the EoF, or negativity, tells us only how entangled a state 
is.  For quantum information purposes, we may like to know what 
{\sl type} of entanglement is present in the system.  For example, 
for distillation protocols, we may want to know if a type of 
qutrit entanglement is present.  This is particularly relevant 
given that some quantum information protocols require entangled 
qudits.  Let us consider what we may discern from 
Eq.~(\ref{eq:rhopt}).


\subsection{Number of Negative Eigenvalues}

The number of negative 
eigenvalues of the partially transposed joint state provides a 
sufficient condition for 
stratification of the pure state entanglement.  

Before addressing this point, recall from 
Sec.~\ref{sec:expid} that given some 
prior knowledge that a bipartite system is in a DPS, one may 
obtain the eigenvalues, i.e. the set $\{b_i\}$, as well as 
$p$ from the spectrum 
of one of the local density operators alone, e.g. from 
$\rho_A=\tr_B(\rho_d)$.  
In what follows, it is assumed that the state is in a DPS and 
that $p$ and $\{b_i\}$ have been determined.  

From Eq.~(\ref{eq:rhopt}), the eigenvalues of the partially 
transposed density operator will be 
\beq
\begin{array}{c}
\Big\{ \{(1-p)\frac{1}{D} + pb_j^2\}_{j=0}^{D_A-1}, 
 \{(1-p)\frac{1}{D} + pb_jb_j^\prime\}_{j<j^\prime}^{D_A-1}, \\
\{(1-p)\frac{1}{D} - pb_jb_j^\prime\}_{j<j^\prime}^{D_A-1}
\Big\}
\end{array}
\eeq
Note that the number of negative eigenvalues is bounded above by 
${D_A \choose 2}$.  For two qubits this means that the maximum 
number of negative eigenvalues is one.  For two qutrits, the 
maximum number of negative eigenvalues is three, etc.  
Note that for a maximally entangled state of two identical 
systems of dimension $D_A$, 
\beq
\Phi_{m} = \frac{1}{\sqrt{D_A}}\sum_{i=0}^{D_A-1}\ket{ii},
\eeq
and symmetry requires that there are either $D_A$ negative 
eigenvalues or none.  
This result supports the conjecture by Han, et al.
\cite{Han/etal:06} that for the maximum number of negative 
eigenvalues for a bipartite entangled mixed state 
is $D_A(D_A-1)/2$.  (Recall $D_A\leq D_B$.)

For example, consider $D=9,$ and $D_A=3=D_B$.  The eigenvalues 
of the partially transposed density operator are
\[
\begin{array}{lll}
\left(\frac{1-p}{9}+pb_1^2\right), \left(\frac{1-p}{9}+pb_2^2\right), 
\left(\frac{1-p}{9}+pb_3^2\right), \\
\left(\frac{1-p}{9}+pb_1b_2\right), \left(\frac{1-p}{9}+pb_1b_3\right), 
\left(\frac{1-p}{9}+pb_2b_3\right),\\
\left(\frac{1-p}{9}-pb_1b_2\right),\left(\frac{1-p}{9}-pb_1b_3\right),
\left(\frac{1-p}{9}-pb_2b_3\right).
\end{array}
\]
By inspection, any of the last three will be negative when 
\[
p > \frac{1}{9b_jb_{j'} +1},
\]
for a given $j,j'$ as is consistent with the general requirement 
that the state be entangled according to the negativity.  However, 
note that if $\dmat{\Psi}{\Psi}$ corresponds to a Bell state, then 
$b_1=\frac{1}{\sqrt{2}} = b_2$ and $b_3 =0$.  This implies that there is 
at most one negative eigenvalue which occurs when $p>2/11$.  
Now consider the maximally entangled two-qutrit state, 
$b_1=b_2=b_3=1/\sqrt{3}$ (or any state locally equivalent to an 
SU(3) singlet).  In this case, when $p>1/4$, all of the last 
three eigenvalues are negative.  Clearly this cannot happen for 
$\dmat{\Psi}{\Psi}$ a two qubit density operator since, at most, 
one eigenvalue is negative.  The difference in the number of 
negative eigenvalues therefore provides 
a sufficient condition for distinguishing two different types of 
entangled states.  
Note that the negativity for the two cases can be the same.  
As a simple example, consider the parameter sets 
1) $p=1/3,b_1=1/\sqrt{2},b_2=1/\sqrt{2},b_3=0$ and 
2) $p=23/72,b_1=1/\sqrt{3},b_2=1/\sqrt{3},b_3=1/\sqrt{3}$.  
Each produces a negativity of $\mathcal{N} = 5/54$.  
It must also be true for any entanglement measure which 
provides only one number to quantify the entanglement, that 
there exists parameters for which the entanglement is the 
same, but the types of entanglement are different.

Since the $p,$ and $b_i$ are measurable quantities, we may 
determine the number of negative eigenvalues.  Alternatively, 
we could determine number of times the coefficients of 
the characteristic polynomial of $\rho_{AB}^{T_B}$ change 
sign. This is equal to the number of positive eigenvalues.  
(See \cite{Byrd/Khaneja:03,Kimura}.)  
Thus the number of negative eigenvalues of the partially 
transposed density operator can be extracted experimentally
and provide a sufficient condition for distinguishing between 
types of entangled states.


\section{Conclusions}
\label{sec:concl}

DPS are simply described in terms of a pure state 
component and a polarization length.  Each of these 
states has a large invariant subspace making it tractable 
to compute in closed form several quantities such as 
distance metrics between states and entanglement between 
subsystems in a joint depolarized state.
Such quantities are useful for determining the 
distinguishability of quantum states and 
the nature of quantum correlations that could be 
used for tasks such as entanglement distillation.  

Aside from their simplicity, there is a 
physical motivation for studying such states: namely, a continuous 
subset of such states corresponds 
to output states from physically allowed depolarization channels.  
Any completely positive map can 
be driven to a depolarization channel by suitable stochastic unitary 
operations, and the strength of the depolarization
is dictated by the magnitude of the identity component of the map.  
In this sense the PDPS correspond to the output 
of a standard form of quantum maps with a pure state input.  
We have described how to 
experimentally measure the parameters of a DPS by measuring 
invariants generated by 
conditions on the coherence vector describing the state.  
Generically, a measurement of all $D$ such invariants 
on an arbitrary quantum state will 
allow for a complete reconstruction of the spectrum of the state.  
However, given prior knowledge that the state is a DPS (for example
by beginning with a pure state, applying an unknown quantum map,
and depolarizing), one can obtain the relevant data by simpler means.
Specifically by measuring two 
quantities $\tr [\rho^2]$ and $\tr[\rho^3]$, 
one obtains the depolarization strength.  
For bipartite systems, measurements of the reduced 
state spectrum then allows for a sufficient measure of 
entanglement between subsystems 
via the negativity.  This requires only $D_A$ measurements 
and is a considerable 
simplification versus tomography on the joint state. 
These measurements can also be used to find the number 
of negative eigenvalues of the partially transposed 
density operator.  This number can be used to 
provide qualitative information about the type, as 
well as amount of entanglement present in the joint state.  
This could, for example, help to distinguish between 
SU(2) and SU(3) singlet states thus providing information 
about the types of interaction between two distant objects.  

We have shown that for 
bipartite systems with composite dimension
$D=D_AD_B$, the negativity of DPS is identically 
zero if $p\leq 1/(D\max(b_jb_{j'})+1)$.  
Yet it is also known that the state is separable if $p<1/(D/2+1)$.   
Do there exist
bound entangled DPS in between?  Verifying the existence of 
bound entangled states 
requires searching in the region of positive partial 
transpose states for states 
which are not separable.  This can be done by 
constructing operators which give 
witness to separability.  Many results have been obtained for low rank states 
\cite{Horodecki:00}, but our case
is maximal rank (because of the presence of the identity component).  
Recently, 
work \cite{Baumgartner:06} has shown the existence of 
optimal separability witnesses for a class of three 
parameter mixed states.  These states are bipartite systems with equal 
dimension composed of the identity mixed with three 
maximally entangled states (locally equivalent to the 
state $\ket{\Phi^+}$).  The authors numerically find 
bound entangled states when two of the parameters are 
nonzero.  It is possible that 
this analysis could also assist in finding, or ruling 
out, bound entangled DPS.


\appendix


\section{Examples}

\label{sec:examples}

The examples of this appendix show some reproductions of known 
results using our simplified methods.  The first provides a canonical 
form for DPS for all two qubit states.  The second example shows 
that the limits for the separability of isotropic states can 
be derived using our methods given that isotropic states are 
a subset of the DPS.


\subsection{Two Qubits}

This section contains an explicit example of two qubits which 
are in DPS form.  The example shows the reduced number of 
parameters--just two relevant parameters for two qubits 
determined up to local unitaries--which are present in a DPS.

If a pure state represented by $\rho$ is acted upon by a depolarizing 
channel and the original $\rho_0$ was a singlet state for two qubits, the 
result of the depolarizing channel is called a Werner state \cite{Werner:89}.  
For a four dimensional Hilbert space, the density matrix has the 
form \cite{Jakob:01,Byrd/Khaneja:03,Kimura}
\beq
\label{eq:22sts}
\rho = \frac{1}{4}\left({\bf 1} 
          + \sqrt{6} \; \vec{n}\cdot \vec{\lambda}\right).
\eeq
This may also be written, for a two-qubit system as 
\beq
\label{eq:22stsex}
\rho = \frac{1}{4}({\bf 1}\otimes {\bf 1} 
        + \vec{S}^A\cdot \vec{\sigma}^A \otimes {\bf 1} 
	+ {\bf 1} \otimes \vec{S}^B \cdot \vec{\sigma}^B
	+ c_{ij} \sigma_i^A \otimes \sigma_j^B).
\eeq
In this expression the constant factor, $\sqrt{6}$ has been absorbed 
into the expansion coefficients $\vec{S}^A, \vec{S}^B$ which represent 
the ``spins'' of the first and second particles respectively and 
$c_{ij}$ which represents the correlations between particle states. 
In Ref.~\cite{Aravind/2sts} an explicit canonical 
form is given for a two qubit pure-state density matrix in 
the Schmidt form:
\bea
\rho &=& \frac{1}{4}({\bf 1}\otimes {\bf 1} + \cos(\Omega) \sigma_z \otimes {\bf 1} 
	+ \cos(\Omega) {\bf 1} \otimes \sigma_z \nonumber \\
	&&
	+ \sin(\Omega) \sigma_x \otimes \sigma_x
	- \sin(\Omega) \sigma_y \otimes \sigma_y
	+ \sigma_z \otimes \sigma_z) \nonumber. \\
\eea
where $0 \leq \Omega \leq \pi/2$. The original 
density matrix is related to this one by a set of local unitary 
transformations. The fact that an explicit canonical form 
has been given for the pure state density matrix of two qubits 
implies that an explicit form of a depolarized pure state density 
matrix can also be given.  

For a DPS the canonical form is given by 
\bea
\rho_d &=& \frac{1}{4}({\bf 1}\otimes {\bf 1} + p\cos(\Omega) \sigma_z \otimes {\bf 1} 
	+ p\cos(\Omega) {\bf 1} \otimes \sigma_z \nonumber \\
	&&
	+ p\sin(\Omega) \sigma_x \otimes \sigma_x
	- p\sin(\Omega) \sigma_y \otimes \sigma_y
	+ p\sigma_z \otimes \sigma_z). \nonumber \\
\eea
The partial transpose (with respect to either subsystem) 
gives the following eigenvalues,
\beq
\begin{array}{lll}
\mu_1&=&\frac{1}{4}(1+p)+\frac{1}{2}p\cos(\Omega);\quad \mu_2=\frac{1}{4}(1+p)-\frac{1}{2}p\cos(\Omega) \\
\mu_3&=&\frac{1}{4}(1-p)+\frac{1}{2}p\sin(\Omega);\quad\mu_4=\frac{1}{4}(1-p)-\frac{1}{2}p\sin(\Omega).
\end{array}
\eeq
The only eigenvalue which could be negative is $\mu_4$.  
However, $\mu_4$ cannot be negative for $p\leq 1/3=1/(D/2+1)$ 
which is the condition derived earlier.  
Otherwise, if $p > 1/3$ we can have entangled states when  
\beq
\sin(\Omega) > \frac{1-p}{2p}.
\eeq
Note that, if $p<0$, the roles of $\mu_2$($\mu_1$) and 
$\mu_4$ ($\mu_3$) are reversed and the same conditions apply 
with the added condition that the original density operator 
is not positive if $p<-1/3$.  

One might refer to these as generalized Werner states since,
unlike the original Werner states, there are two variable 
parameters. One describes the magnitude of the coherence vector and 
the other describes the entanglement of the pure state to which 
the DPS corresponds.  


\subsection{Isotropic States}

We can verify an entanglement property of isotropic states with this result.  
Isotropic states are defined over bipartite states of equal dimension, 
$(D_A=D_B)$ by a single parameter $F$
\[
\rho_d=\frac{1-F}{D_A^2-1}({\bf 1}_{AB}-\ket{\Phi^+}_{AB}\bra{\Phi^+})+F\ket{\Phi^+}_{AB}\bra{\Phi^+}
\]
where $\ket{\Phi^+}_{AB}=\frac{1}{\sqrt{D_A}}\sum_{j=0}^{D_A-1} \ket{j}_A\ket{j}_B$.  
This class of states parameterizes depolarized maximally entangled states where 
$p=(D_A^2F-1)/(D_A^2-1)$.  The entanglement properties of isotropic states have 
been studied before and it has been shown \cite{Horodecki:99} that $\rho_F$ 
is separable iff $0\leq F\leq 1/D_A$, or
$-1/(D_A^2-1)\leq p \leq 1/(D_A+1)$.  This is consistent with the above 
result as all pairs of Schmidt coefficients for $\rho_F$
 have the value $b_j b_{j'}=1/D_A$ which means that for 
$p>1/(D_A+1)$, the isotropic states are entangled.  Consequently,
there are no bound entangled isotropic states.


\section*{ACKNOWLEDGMENTS}

MSB gratefully acknowledges 
C. Allen Bishop and  Todd Tilma for helpful discussions.  
This material is based upon work supported by the 
National Science Foundation under Grant No. 0545798 to MSB. 
GKB received support from the Austrian Science Foundation.



\begin{thebibliography}{59}
\bibitem{Blochnote}
\bibinfo{note}{{For $D=2$, any pair of orthogonal pure states are represented
  as antipodal points on the Bloch sphere with opposite coherence vectors
  $\vec{n},-\vec{n}$. But then $-\vec{n}=(-\vec{n})\star (-\vec{n})
  =\vec{n}\star \vec{n}=\vec{n}$ implies $\vec{n}\star \vec{n}=0$.}}
\bibitem{diagnote}
\bibinfo{note}{{If $D_A=D_B$ then this is a diagonalization procedure.
  Otherwise, we can append rows/columns of zeros so that $a_{i,\mu}$ is a
  square matrix and again say that $a$ is diagonalized by $U$ and $V$. This is
  the singular value decomposition.}}
\end{thebibliography}
\end{document}